\documentclass[prb,twocolumn,showpacs,superscriptaddress]{revtex4}

\usepackage[dvips]{graphicx}
\usepackage{dcolumn}% Align table columns on decimal point
\usepackage{bm}% bold math
\usepackage{amsmath}
\usepackage{ulem}

\begin{document}

%abbreviations
\newcommand{\1}{{\bf \scriptstyle 1}\!\!{1}}
\newcommand{\p}{\partial}
\newcommand{\D}{^{\dagger}}
\newcommand{\bx}{{\bf x}}
\newcommand{\bk}{{\bf k}}
\newcommand{\bv}{{\bf v}}
\newcommand{\bp}{{\bf p}}
\newcommand{\bu}{{\bf u}}
\newcommand{\bA}{{\bf A}}
\newcommand{\bB}{{\bf B}}
\newcommand{\bD}{{\bf D}}
\newcommand{\bK}{{\bf K}}
\newcommand{\bL}{{\bf L}}
\newcommand{\bP}{{\bf P}}
\newcommand{\bQ}{{\bf Q}}
\newcommand{\bS}{{\bf S}}
\newcommand{\bH}{{\bf H}}
\newcommand{\bg}{{\bf g}}
\newcommand{\balpha}{\mbox{\boldmath $\alpha$}}
\newcommand{\bsigma}{\mbox{\boldmath $\sigma$}}
\newcommand{\bSigma}{\mbox{\boldmath $\Sigma$}}
\newcommand{\bOmega}{\mbox{\boldmath $\Omega$}}
\newcommand{\bpi}{\mbox{\boldmath $\pi$}}
\newcommand{\bphi}{\mbox{\boldmath $\phi$}}
\newcommand{\bnabla}{\mbox{\boldmath $\nabla$}}
\newcommand{\bmu}{\mbox{\boldmath $\mu$}}
\newcommand{\bepsilon}{\mbox{\boldmath $\epsilon$}}

\newcommand{\iLambda}{{\it \Lambda}}
\newcommand{\cA}{{\cal A}}
\newcommand{\cD}{{\cal D}}
\newcommand{\cL}{{\cal L}}
\newcommand{\cH}{{\cal H}}
\newcommand{\cI}{{\cal I}}
\newcommand{\cM}{{\cal M}}
\newcommand{\cO}{{\cal O}}
\newcommand{\cR}{{\cal R}}
\newcommand{\cU}{{\cal U}}
\newcommand{\cT}{{\cal T}}

\newcommand{\be}{\begin{equation}}
\newcommand{\ee}{\end{equation}}
\newcommand{\bea}{\begin{eqnarray}}
\newcommand{\eea}{\end{eqnarray}}
\newcommand{\beqa}{\begin{eqnarray*}}
\newcommand{\eeqa}{\end{eqnarray*}}
\newcommand{\nn}{\nonumber}
\newcommand{\DD}{\displaystyle}

\newcommand{\ba}{\left[\begin{array}{c}}
\newcommand{\baa}{\left[\begin{array}{cc}}
\newcommand{\baaa}{\left[\begin{array}{ccc}}
\newcommand{\baaaa}{\left[\begin{array}{cccc}}
\newcommand{\ea}{\end{array}\right]}

\title{Phonon Bottleneck Effect Leads to Observation of Quantum Tunneling\\
of the Magnetization and Butterfly Hysteresis Loops in
(Et$_{4}$N)$_{3}$Fe$_{2}$F$_{9}$}

\author{Ralph Schenker}
\email{ralph.schenker@iac.unibe.ch}
\affiliation{Department for Chemistry and
Biochemistry, University of Bern, Freiestrasse 3, 3000 Bern 9, Switzerland}

\author{Michael N.~Leuenberger}
\altaffiliation{Present address: Department of Physics, University of California San
Diego, 9500 Gilman Drive, La Jolla, CA 92093-0360}\email{mleuenbe@physics.ucsd.edu}
\affiliation{Department of Physics and Astronomy, University of Basel,
Klingelbergstrasse 82, 4056 Basel, Switzerland}

\author{Gr\'egory Chaboussant}
\altaffiliation{Present address: Laboratoire L\'{e}on Brillouin (LLB-CNRS-CEA), CEA
Saclay, 91191 Gif-sur-Yvette Cedex, France}\email{chabouss@llb.saclay.cea.fr}
\affiliation{Department for Chemistry and Biochemistry, University of Bern,
Freiestrasse 3, 3000 Bern 9, Switzerland}

\author{Daniel Loss}
\email{daniel.loss@unibas.ch} \affiliation{Department of Physics and Astronomy,
University of Basel, Klingelbergstrasse 82, 4056 Basel, Switzerland}

\author{Hans U. G\"{u}del}
\email{hans-ulrich.guedel@iac.unibe.ch} \affiliation{Department for Chemistry and
Biochemistry, University of Bern, Freiestrasse 3, 3000 Bern 9, Switzerland}

\begin{abstract}
A detailed investigation of the unusual dynamics of the magnetization of
(Et$_4$N)$_3$Fe$_2$F$_9$ (Fe$_2$), containing isolated {[Fe$_2$F$_9$]}$^{3-}$ dimers, is
presented and discussed. Fe$_2$ possesses an $S=5$ ground state with an energy barrier of
2.40 K due to an axial anisotropy. Poor thermal contact between sample and bath leads to
a phonon bottleneck situation, giving rise to butterfly-shaped hysteresis loops below 5 K
concomitant with slow decay of the magnetization for magnetic fields $H_z$ applied along
the Fe--Fe axis. The butterfly curves are reproduced using a microscopic model based on
the interaction of the spins with resonant phonons. The phonon bottleneck allows for the
observation of resonant quantum tunneling of the magnetization at 1.8 K, far above the
blocking temperature for spin-phonon relaxation. The latter relaxation is probed by AC
magnetic susceptibility experiments at various temperatures and bias fields $H_{\rm DC}$.
At $H_{\rm DC}=0$, no out-of-phase signal is detected, indicating that at $T\geq1.8$ K
Fe$_2$ does not behave as a single-molecule magnet. At $H_{\rm DC}=1$ kG, relaxation is
observed, occurring over the barrier of the thermally accessible $S=4$ first excited
state that forms a combined system with the $S=5$ state.
\end{abstract}


\maketitle

% INTRODUCTION
\section{Introduction}
Among the most fascinating aspects of magnetism in recent years has been the discovery
that the magnetic moment of individual molecules can give rise to magnetic hysteresis
phenomena. Such systems lie at the interface between classical and quantum-mechanical
regimes, and are thus of great fundamental as well as practical interest regarding
future applications related to quantum computing.\cite{Leuenberger2001}
\par
Molecular clusters such as
Mn$_{12}$,\cite{Sessoli1993nature,Thomas1996,Friedman1996,Hernandez1997}
Fe$_8$,\cite{Sangregorio1997} and Mn$_4$\cite{Aubin1998} have been found to exhibit
hysteresis curves characterized by a remanence. They have been dubbed single-molecule
magnets (SMMs) as each molecule behaves as a single-domain nanomagnetic particle. These
high-spin systems possess an energy barrier $\Delta=|D_S|S^{2}$ for reversal of the
direction of the magnetic moment (``spin"), arising from the combination of a large
ground state total spin quantum number $S$ with a uniaxial anisotropy (negative axial
zero--field splitting parameter $D_S$). At $kT\ll\Delta$ the magnetization can be
blocked and relaxes very slowly, giving rise to the hysteresis. Many SMMs such as
Mn$_{12}$\cite{Thomas1996,Friedman1996} and Fe$_8$\cite{Sangregorio1997} exhibit
quantum tunneling of the magnetization (QTM) through the energy barrier.
\par
Alternatively, a different type of hysteresis lacking a remanence has been observed for
[Fe(salen)Cl]$_2$\cite{Shapira1999} and more recently for the $S=1/2$ spin cluster
V$_{15}$\cite{Chiorescu2000} as well as the antiferromagnetically coupled ($S=0$)
ferric wheels Fe$_6$\cite{Waldmann2002} and Fe$_{12}$.\cite{Inagaki2003} In these
low-spin systems the ground state does not possess a magneto-structural energy barrier
for the reversal of the total spin.\cite{Neelvector} Instead, the hysteresis is created
by dynamically driving the spin system out of thermal equilibrium due to insufficient
thermal contact of the sample with the heat bath, thereby causing a phonon bottleneck
(PB) effect. It possesses a characteristic butterfly shape that arises from a fast spin
reversal at the field value where the anticrossing of the two lowest-energy levels
occurs.\cite{Chiorescu2000,Waldmann2002} Generally, the magnitude of the anticrossing
level splitting determines the quantum dynamics of the spins. In the low-spin systems,
this splitting is relatively large ($\geq$10$^{-3}$ K), allowing for fast spin reversal
yielding the butterfly shape. On the other hand, in high-spin systems the tunnel
splitting between the lowest levels of typically $\leq10^{-6}$ K renders QTM too slow
for allowing a butterfly-shaped hysteresis. Therefore it was previously argued that
butterfly hystereses are not expected for high-spin systems.\cite{Chiorescu2000}
Nevertheless, recently for the SMM Mn$_9$ a butterfly hysteresis caused by the PB
effect was reported above the blocking temperature for SMM behavior, but no analysis
was offered.\cite{Boscovic2002IC} Very little is known to date about the physics of
this phenomenon in spin systems with an energy barrier. However, such information is of
particular interest as the presence of the barrier principally enables these clusters
to behave as single-molecule magnets.
\par
The tri-$\mu$-fluoro bridged [Fe$_{2}$F$_{9}$]$^{3-}$ dimer molecule in the compound
(Et$_{4}$N)$_{3}$Fe$_{2}$F$_{9}$ (Et$_{4}$N = tetraethylammonium), abbreviated Fe$_2$,
possesses an $S=5$ ground state with an energy barrier of 2.40 K for spin reversal
(Fig.~\ref{figBarrier}).\cite{Schenker2001IC} As reported
previously,\cite{Schenker2002} Fe$_2$ exhibits slow relaxation of the magnetization and
magnetic hysteresis below 5 K. The hysteresis adopts a butterfly shape tied up at
$H_z=0$ despite the presence of the energy barrier. The observation of QTM prompted us
to ascribe these phenomena to the unusually slow relaxation of a single-molecule
magnet.\cite{Schenker2002} However, meanwhile additional data have provided compelling
evidence that the observed relaxation depends on the thermal insulation of the sample.
\par
Here we present a detailed study on the magnetic properties of Fe$_2$. The earlier
findings are reinterpreted in terms of the phonon bottleneck. A microscopic model based
on the rapid absorption/emission of resonant phonons is developed that allows for an
accurate reproduction of the observed hysteresis curves. Finally, alternating-current
(AC) magnetic susceptibility data reflecting the fast relaxation behavior of Fe$_2$
unaffected by the phonon bottleneck effect are presented and discussed.

%%%%%%%%%%%%%%%%%%%%%%%%%%%%%%%%%%%%%%%%%%%%%%%%%%%%%%%%%%%%%%%%%%%%%%%%%%%%%%%%%%%%%%%%%%%%%%%%%%%%%%%%%%%%%%%%%%%%%
\begin{figure}[!t]
\includegraphics[width=65mm]{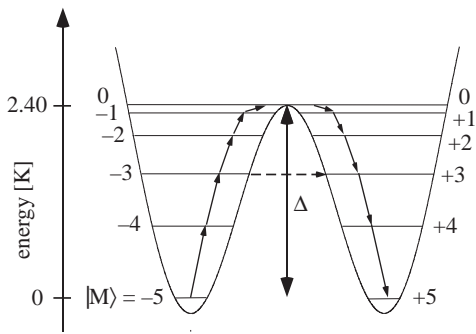}
\caption{Double-well potential of the $S=5$ ground state of Fe$_2$ with the energy
barrier $\Delta=|D_{S}S^2|=2.40$ K. Full and broken arrows indicate spin-phonon and QTM
transitions, respectively.} \label{figBarrier}
\end{figure}
%%%%%%%%%%%%%%%%%%%%%%%%%%%%%%%%%%%%%%%%%%%%%%%%%%%%%%%%%%%%%%%%%%%%%%%%%%%%%%%%%%%%%%%%%%%%%%%%%%%%%%%%%%%%%%%%%%%%%

% EXPERIMENTAL SECTION
\section{Experimental}\label{SecExperimental}
Needle-shaped single crystals of (Et$_{4}$N)$_{3}$Fe$_{2}$F$_{9}$ (Fe$_2$) up to 5 mm
in length were prepared according to Ref.~\onlinecite{Schenker2001IC}. The dimer
symmetry is exactly $C_{3h}$ with the threefold axis lying parallel to the hexagonal
crystal axis {\it c}.\cite{Schenker2001} The intradimer Fe$^{\ldots}$Fe distance is
2.907 {\AA}. The dimer molecules are well separated from each other, leading to
interdimer Fe$^{\ldots}$Fe distances of at least 8 {\AA}, thus making interdimer
interactions extremely inefficient.
\par
Samples for magnetic experiments were prepared as follows: To prevent hydrolyzation,
samples were prepared in a drybox under a nitrogen atmosphere. All the crystals were
perfectly transparent, indicating that no hydrolysis occurred. A single crystal of 2.46
mg (sample {\bf A}) was sealed in a glass tube under N$_2$ in $\bH\parallel c$ ($H_z$)
orientation. At liquid helium temperatures, the N$_2$ is frozen out and its vapor
pressure inside the glass tube becomes very low, leading to a poor heat contact of the
crystal with its surroundings. Alternatively, three single crystals (total mass 48 mg)
were embedded in Apiezon grease in a dried gelatine capsule to yield a sample ({\bf B})
with improved heat contact. Alternating-current (AC) magnetic susceptibility
measurements were performed both on sample {\bf B} and on a polycrystalline powder
sample ({\bf C}) of 25 mg made of crushed single crystals, sealed in a dried gelatine
capsule.
\par
Magnetic measurements were carried out using a Quantum Design MPMS-XL-5 Superconducting
Quantum Interference Device (SQUID) magnetometer equipped with a 5 T magnet. The
oscillation frequency of the AC field was varied between 10 and 1488 Hz at an AC field
strength of 1 G. All data were corrected for the diamagnetism of the atoms using Pascal's
constants.

% RESULTS
\section{Results}

% Relaxation of Magnetization
\subsection{Relaxation of the Magnetization}\label{secMagnetrelax}
Magnetization relaxation experiments were performed using sample {\bf A} in
$\bH\parallel c$ orientation at fixed values of $H_z$ and bath temperature $T$, after
saturation at 30 kG and subsequent quick ($\approx150$ G/s) reduction to the indicated
$H_z$ value. Relaxation times $\tau_2$ were extracted from least-squares fits of a
single-exponential law to the relaxation curves for various values of $H_z$. The
resulting field dependence of $\tau_2$ at 1.8 K (Fig.~\ref{figRelaxationDips}) reveals
that $\tau_2$ varies between 240 s at $H_z=0$ and 870 s at 3.3 kG. Importantly, 3
distinctive dips are observed at $H_z=0$, 1.1, and 2.7 kG.
%
%%%%%%%%%%%%%%%%%%%%%%%%%%%%%%%%%%%%%%%%%%%%%%%%%%%%%%%%%%%%%%%%%%%%%%%%%%%%%%%%%%%%%%%%%%%%%%%%%%%%%%%%%%%%%%%%%%%%%%
\begin{figure}[!t]
\includegraphics[width=70mm]{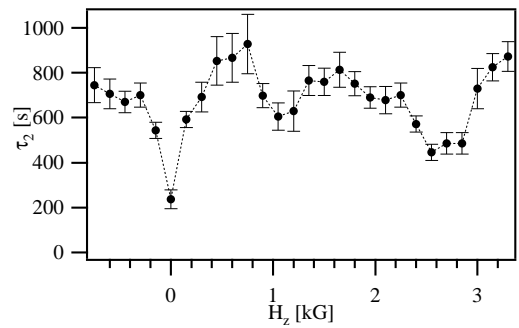}
\caption{Dependence of the relaxation time $\tau_2$ on $H_z$ at $T=1.8$ K with error
(3$\sigma$). The dotted line is a guide to the eyes.} \label{figRelaxationDips}
\end{figure}
%%%%%%%%%%%%%%%%%%%%%%%%%%%%%%%%%%%%%%%%%%%%%%%%%%%%%%%%%%%%%%%%%%%%%%%%%%%%%%%%%%%%%%%%%%%%%%%%%%%%%%%%%%%%%%%%%%%%%%%%

% Isothermal Magnetization: Hysteresis
\subsection{Butterfly Hysteresis Curves}\label{secButterfly}
Magnetization measurements at fixed $T$ were performed in $\bH\parallel c$ ($H_z$)
orientation by varying $H_z$ from +50 kG to --50 kG and back to +50 kG. For the crystal
embedded in Apiezon grease (sample {\bf B}), the magnetization follows the thermal
equilibrium curve (asterisks in Fig.~\ref{figFullhysteresis}). Alternatively, for a
crystal mounted in the glass tube (sample {\bf A}), it features a butterfly-shaped
hysteresis loop characterized by two distinct features: (i) the magnetization at
$H_z=0$ is zero, and (ii) near $H_z=0$ the magnitudes of both the magnetization
detected while approaching $H_z=0$ and the one while receding from it are larger than
the thermal equilibrium value (inset of Fig.~\ref{figFullhysteresis}). As this
butterfly hysteresis is perfectly centrosymmetric, further data in
Fig.~\ref{figHysteresisdependence} show only the positive wings. The hysteresis effect
decreases with increasing $T$, vanishing above 5 K
(Fig.~\ref{figHysteresisdependence}). Reduction of the sweeping rate from
$\Gamma\approx3$ G/s to $\approx$1 G/s~\cite{sweepingrate_fnote} narrows the breadth of
the wings (Fig. 2 in Ref.~\onlinecite{Schenker2002}), suggesting that for a
substantially smaller value of $\Gamma$ the magnetization would follow the thermal
equilibrium curve (dashed line in Fig.~\ref{figFullhysteresis}).
%
%%%%%%%%%%%%%%%%%%%%%%%%%%%%%%%%%%%%%%%%%%%%%%%%%%%%%%%%%%%%%%%%%%%%%%%%%%%%%%%%%%%%%%%%%%%%%%%%%%%%%%%%%%%%%%%%%%%%%
\begin{figure}[!t]
\includegraphics[width=69mm]{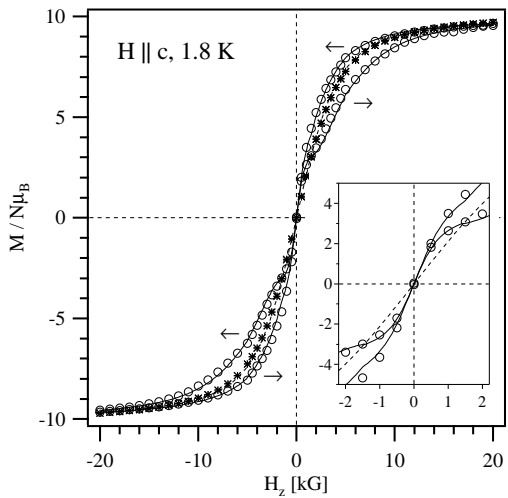}
\caption{Hysteresis of the magnetization of Fe$_2$ single crystals (open circles:
sample {\bf A}, asterisks: sample {\bf B}) at 1.8 K obtained with $\Gamma\approx3$ G/s.
The arrows indicate the direction of the measurement. The solid lines are calculated
using the resonant-phonon model (Sec.~\ref{secAvalanche}). The dashed line represents
the thermal equilibrium curve calculated with $J=-2.23$ K, $D=-0.215$ K, and
$g=2.00$.\cite{Schenker2001IC} The inset highlights the region close to $H_z=0$.}
\label{figFullhysteresis}
\end{figure}
%%%%%%%%%%%%%%%%%%%%%%%%%%%%%%%%%%%%%%%%%%%%%%%%%%%%%%%%%%%%%%%%%%%%%%%%%%%%%%%%%%%%%%%%%%%%%%%%%%%%%%%%%%%%%%%%%%%%%%
%
%%%%%%%%%%%%%%%%%%%%%%%%%%%%%%%%%%%%%%%%%%%%%%%%%%%%%%%%%%%%%%%%%%%%%%%%%%%%%%%%%%%%%%%%%%%%%%%%%%%%%%%%%%%%%%%%%%%%%
\begin{figure}[ht]
\includegraphics[width=69mm]{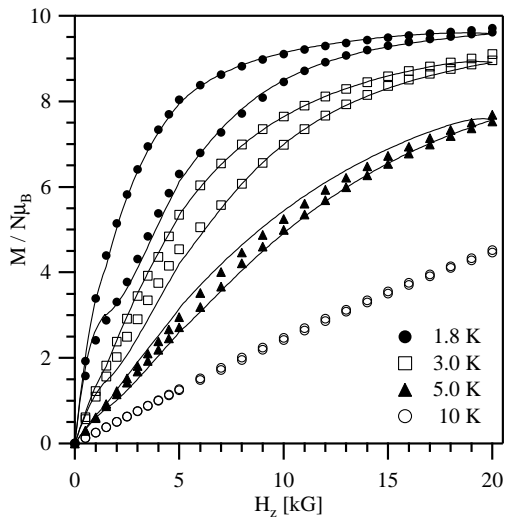}
\caption{Positive wings of the butterfly hysteresis of the magnetization of a single
crystal (sample {\bf A}) for $\bH\parallel c$ orientation at different temperatures as
indicated with $\Gamma\approx3$ G/s. The solid lines are calculated using the
resonant-phonon model (Sec.~\ref{secAvalanche}).} \label{figHysteresisdependence}
\end{figure}
%%%%%%%%%%%%%%%%%%%%%%%%%%%%%%%%%%%%%%%%%%%%%%%%%%%%%%%%%%%%%%%%%%%%%%%%%%%%%%%%%%%%%%%%%%%%%%%%%%%%%%%%%%%%%%%%%%%%%
\par
A drastic change in the magnetization curve of sample {\bf A} occurs upon a very fast
sweep of the magnetic field. Fig.~\ref{figAvalanche} shows magnetization data obtained
as follows: after saturation at +30 kG, the field was quickly ($\Gamma\approx150$ G/s)
reduced to the given values of $H_z$ including negative ones, and the magnetization was
measured within 50 s after reaching $H_z$. During such a fast sweep, the magnetization
essentially switches from positive to negative saturation upon crossing $H_z=0$.

%%%%%%%%%%%%%%%%%%%%%%%%%%%%%%%%%%%%%%%%%%%%%%%%%%%%%%%%%%%%%%%%%%%%%%%%%%%%%%%%%%%%%%%%%%%%%%%%%%%%%%%%%%%%%%%%%%%%%
\begin{figure}[!t]
\includegraphics[width=69mm]{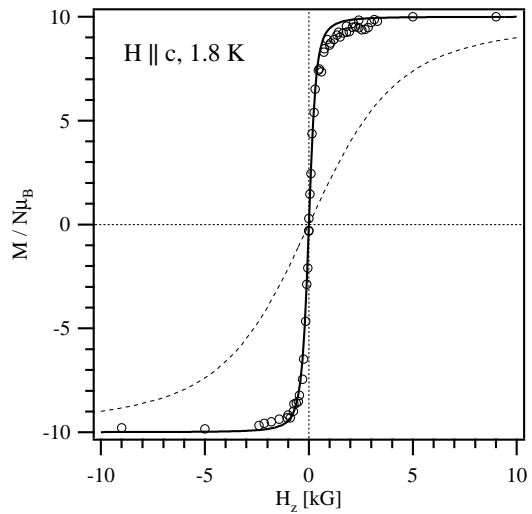}
\caption{Magnetization data (sample {\bf A}) at 1.8 K obtained as follows: after
saturation at +30 kG the field was quickly ($\Gamma\approx150$ G/s) reduced to the
given $H_z$ values including negative ones, and the magnetization was measured within
50 s after reaching $H_z$. The solid line is the adiabatic curve calculated using the
resonant-phonon model described in the text with $H_{{\rm eff},x}=350$ G, while the
dashed line represents the thermal equilibrium curve.} \label{figAvalanche}
\end{figure}
%%%%%%%%%%%%%%%%%%%%%%%%%%%%%%%%%%%%%%%%%%%%%%%%%%%%%%%%%%%%%%%%%%%%%%%%%%%%%%%%%%%%%%%%%%%%%%%%%%%%%%%%%%%%%%%%%%%%%%

%%%%%%%%%%%%%%%%%%%%%%%%%%%%%%%%%%%%%%%%%%%%%%%%%%%%%%%%%%%%%%%%%%%%%%%%%%%%%%%%%%%%%%%%%%%%%%%%%%%%%%%%%%%%%%%%%%%%%
\begin{figure}[htb]
\includegraphics[width=75mm]{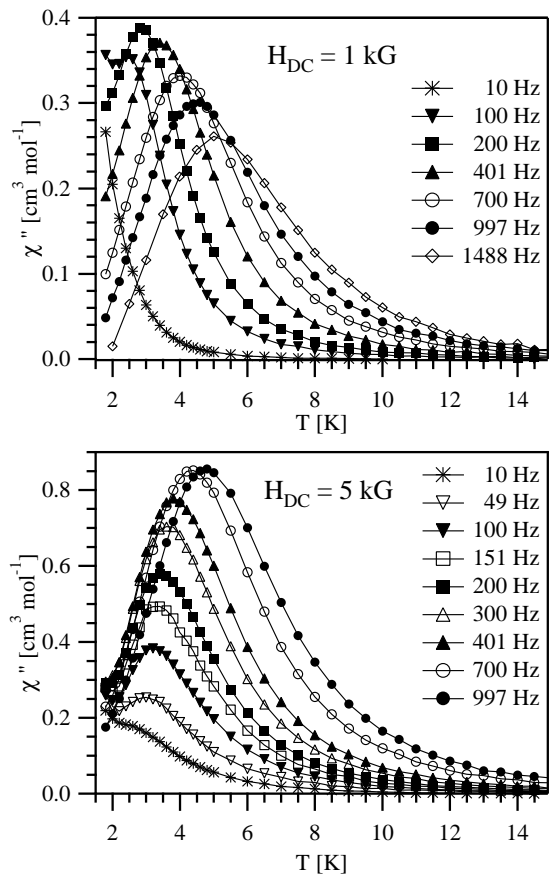}
\caption{Out-of-phase AC susceptibility $\chi$" vs $T$ at $H_{\rm DC}=1$ kG (top) and 5
kG (bottom) for a polycrystalline sample ({\bf C}) of Fe$_2$ in a 1 G AC field
oscillating with the indicated frequencies $\nu$.} \label{figACfrequency}
\end{figure}
%%%%%%%%%%%%%%%%%%%%%%%%%%%%%%%%%%%%%%%%%%%%%%%%%%%%%%%%%%%%%%%%%%%%%%%%%%%%%%%%%%%%%%%%%%%%%%%%%%%%%%%%%%%%%%%%%%%%%

% AC susceptibility
\subsection{AC Magnetic Susceptibility}\label{secACresults}
To complement our studies on the magnetization relaxation in Fe$_2$,
alternating-current (AC) susceptibility experiments were performed on both the
polycrystalline sample {\bf C} and the single crystals in $\bH_{\rm
AC}\parallel\bH_{\rm DC}\parallel c$ orientation embedded in Apiezon grease (sample
{\bf B}). $H_{\rm DC}$ denotes a static bias field and for sample {\bf B} corresponds
to $H_z$. With respect to relaxation, no significant differences were found between the
two samples.\cite{ACXtalpowder_fnote} Experiments were conducted within a temperature
range $1.8\leq T\leq15$ K with an amplitude of $H_{\rm AC}=1$ G and oscillation
frequencies $10\leq\nu\leq1488$ Hz. Importantly, an out-of-phase susceptibility
($\chi$") was observed only upon application of an additional static magnetic field
$H_{\rm DC}$.
\par
Figure~\ref{figACfrequency} shows the frequency dependence of $\chi$" vs $T$ at
selected static fields of 1 kG and 5 kG using the polycrystalline sample. As $\nu$ is
reduced from 1488 Hz to 100 Hz, at 1 kG the peak maximum $T_{\rm max}$ shifts from 5.2
K to 2.45 K, and further reduction to 10 Hz causes it to shift below 1.8 K. At 5 kG the
situation is similar. In both cases $T_{\rm max}$ increases with $\nu$, from which the
temperature dependence of the relaxation rate $1/\tau=2\pi\nu$ is derived, shown in
Fig.~\ref{figACarrhenius} as Arrhenius plots of ln$(1/\tau)$ versus $1/T_{\rm max}$. At
$H_{\rm DC}=1$ kG the data follow the Arrhenius law
\begin{equation}
\frac{1}{\tau}=\frac{1}{\tau_0}\cdot e^{-\Delta E/kT}\quad ,\label{eqnArrhenius}
\end{equation}
where $1/\tau_0=1.0\times10^{5}$ s$^{-1}$ is the intrinsic relaxation rate and $\Delta
E=12.6\pm0.2$ K the kinetic energy barrier. At $H_{\rm DC}=5$ kG, however, the
temperature dependence of $1/\tau$ is linear (inset of Fig.~\ref{figACarrhenius}).
\par
Interestingly, for higher frequencies, $\chi$" is considerably larger at 5 kG than at 1
kG. Indeed, we observed that for $\nu=997$ Hz, $\chi$" increases linearly with $H_{\rm
DC}$ below 2.5 kG and appears to level off at higher fields (not shown).

%%%%%%%%%%%%%%%%%%%%%%%%%%%%%%%%%%%%%%%%%%%%%%%%%%%%%%%%%%%%%%%%%%%%%%%%%%%%%%%%%%%%%%%%%%%%%%%%%%%%%%%%%%%%%%%%%%%%%
\begin{figure}[!t]
\includegraphics[width=75mm]{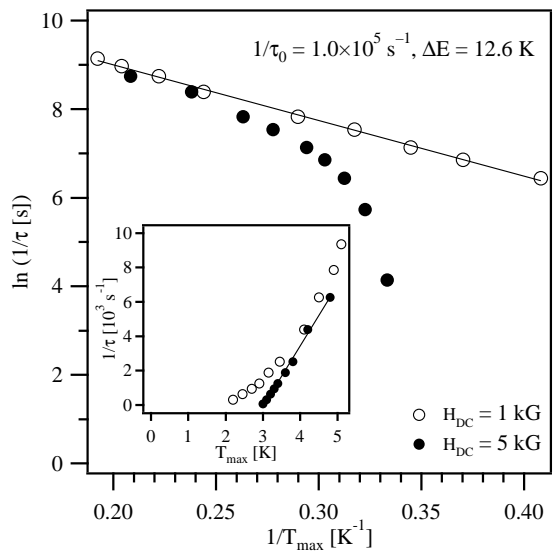}
\caption{Temperature dependence of the AC $\chi$" relaxation rate $1/\tau=2\pi\nu$
shown as an Arrhenius plot (ln $1/\tau$ {\it vs} 1/$T_{\rm max}$) as extracted from the
data in Fig.~\ref{figACfrequency}. The solid line represents a least--squares fit of
the $H_{\rm DC}=1$ kG data to the Arrhenius law (Eq.~\ref{eqnArrhenius}), yielding the
prefactor $1/\tau_0$ and the kinetic energy barrier $\Delta E$ as indicated. The inset
shows the same data on a linear scale. Note the linear dependence of $1/\tau$ on
$T_{\rm max}$ for the $H_{\rm DC}=5$ kG data.} \label{figACarrhenius}
\end{figure}
%%%%%%%%%%%%%%%%%%%%%%%%%%%%%%%%%%%%%%%%%%%%%%%%%%%%%%%%%%%%%%%%%%%%%%%%%%%%%%%%%%%%%%%%%%%%%%%%%%%%%%%%%%%%%%%%%%%%%

% ANALYSIS
\section{Analysis and Discussion}
\label{analysis}
% Double Well Potential
\subsection{Phonon Bottleneck Effect}\label{secPBE}
In $H_z$ orientation, Fe$_2$ single crystals in a glass tube (sample {\bf A}) exhibit
slow magnetization relaxation on the timescale of 10$^{2}-10^{3}$ s, whereas crystals
embedded in Apiezon grease (sample {\bf B}) do not. Obviously this relaxation depends
on the degree of thermal insulation of the sample. Poor thermal contact between the
crystal and the heat bath (as for sample {\bf A}) is well known to lead to the
macroscopic observation of the phonon bottleneck (PB).\cite{Vleck1941,Bindilatti1991}
Also, the butterfly hysteresis loops observed for Fe$_2$ closely resemble those
recently reported for the spin clusters V$_{15}$,\cite{Chiorescu2000}
Fe$_6$,\cite{Waldmann2002} and Fe$_{12}$\cite{Inagaki2003} that have been attributed to
this effect. Thus, it is the phonon bottleneck that causes the observed slow relaxation
in Fe$_2$.
\par
The energy exchange between the spin system and the bath occurs via the phonons in the
crystal. The PB describes the fact that at low temperatures the number of spins is much
larger than the number of available phonons. Thermodynamically speaking, the heat
capacity of the spins $C_s$ far exceeds that of the phonons $C_p$:\cite{Abragam1970}
\begin{equation}
b=\frac{C_s+C_p}{C_p}\approx\frac{C_s}{C_p}\gg1\quad,  \label{eqndef_b}
\end{equation}
such that $b\approx10^{4}-10^{6}$.\cite{Abragam1970} The observable relaxation time
is\cite{Abragam1970}
\begin{equation}
\tau_2=\tau_{\rm sp}+b\tau_{\rm pb}\quad , \label{eqndef_t2}
\end{equation}
with $\tau_{\rm sp}$ and $\tau_{\rm pb}$ denoting the spin-phonon and phonon-bath
relaxation times, respectively. Importantly, upon poor thermal contact between the
phonons and the bath, $\tau_{\rm pb}$ becomes very large such that $\tau_2\approx
b\tau_{\rm pb}$, thus rendering the bottleneck macroscopically observable. In a PB
situation, the initial rapid energy transfer from the spins to the phonons quickly
heats the latter to the temperature of the former, such that these phonons become in
resonance with the energy differences between individual $\left|M\right>$
levels.\cite{Abragam1970} Thus, the spins and these resonant phonons form a single
coupled system that can only very slowly exchange energy with the bath, leading to the
observed magnetization relaxation.

% Quantum tunneling
\subsection{Quantum Tunneling of the Magnetization in ${\bf Fe_2}$}\label{secQuantumtunneling}
A very important result of our studies is the observation of dips in the field
dependence of $\tau_2$ (Fig.~\ref{figRelaxationDips}). Such dips are often observed in
single-molecule magnets such as Mn$_{12}$,\cite{Thomas1996,Friedman1996} and
interpreted as fingerprints for resonant QTM at the respective field values, which
lowers the relaxation time. In the case of Fe$_2$, the dependence of $\tau_2$ on $H_z$
appears very intriguing as the slow relaxation arises from the weak thermal contact
between the sample and the bath, which hardly depends on $H_z$. However, since a
tunneling pathway represents a bypass, QTM occurs in parallel to thermal relaxation
involving phonons.\cite{Leuenberger2000} Thus, its presence can significantly alter the
observed relaxation time $\tau_2$ even if the latter is dominated by the phonon-bath
relaxation. Indeed, recently Chiorescu et al. reported magnetization relaxation data on
the V$_{15}$ spin cluster under virtually adiabatic conditions analogous to our
situation for Fe$_2$.\cite{Chiorescu2003} The authors found that around $H_z=0$ the
measured phonon-bath relaxation time $\tau_2$ dropped by a factor of $2-3$ due to the
level anticrossing.
\par
In the $C_{3h}$ dimer symmetry of Fe$_2$,\cite{Schenker2001} the full anisotropy
Hamiltonian including higher--order terms reads\cite{Abragam1970}
\begin{equation}
\cH_{\rm{ani}}=D_{S}\left[\hat{S}_{z}^{2}-\frac{1}{3}S(S+1)\right]+ B_4^0\hat{O}_4^0
 + B_6^0\hat{O}_6^0 + B_6^6\hat{O}_6^6 ,
\label{eqnAnisotropy}
\end{equation}
where
$\hat{O}_4^0=35\hat{S}_z^4-30S(S+1)\hat{S}_z^2+25\hat{S}_z^2-6S(S+1)+3\hat{S}^2(S+1)^2$,
$\hat{O}_6^0 =
231\hat{S}_z^6-315S(S+1)\hat{S}_{z}^{4}+735\hat{S}_{z}^{4}+105S^2(S+1)^2\hat{S}_{z}^{2}
 - 525S(S+1)\hat{S}_{z}^{2}+294\hat{S}_{z}^{2}-5S^3(S+1)^3+40S^2(S+1)^2-60S(S+1)$, and
$\hat{O}_6^6=(S_+^6+S_-^6)/2$. The $B_6^6\hat{O}_6^6$ term mixes wavefunctions with
$\Delta M=\pm6$, allowing for resonant tunneling at $H_z=0$ between the $|M\rangle$
levels --3/+3 (Fig.~\ref{figBarrier}) and at applied fields when the --2/+4 and --1/+5
levels cross. For $B_4^0=B_6^0=0$, the fields $H_z^{MM'}$ at which resonant tunneling
is expected for the $S=5$ state in Fe$_2$ are given by\cite{Leuenberger2000}
\begin{equation}
H_z^{MM'} = -\frac{nD_5}{g\mu_{B}} = n\cdot 714\quad G\quad ,
\end{equation}
where $n=M+M'$ is an integer ranging from 0 to 9. Resonant tunneling may therefore
occur for $n=0$, 2, and 4 at field values of $H_z=0$, 1.4, and 2.85 kG, corresponding
to the $|M\rangle$ level crossings --3/+3 (Fig.~\ref{figBarrier}), --2/+4, and --1/+5,
respectively. Therefore we can expect three dips in the field dependence of $\tau_2$,
in agreement with the experimental data in Fig.~\ref{figRelaxationDips}. The dips are
centered at 0, $\approx$1.1, and $\approx$2.7 kG instead of 0, 1.4, and 2.85 kG,
suggesting nonzero values for $B_4^0$ and $B_6^0$ in Eq.~(\ref{eqnAnisotropy}). Thus
the observed dips in Fig.~\ref{figRelaxationDips} are consistent with quantum tunneling
processes in Fe$_2$ at 1.8 K. However, QTM does not appear to be dominant in Fe$_2$ as
the drop of $\tau_2$ due to QTM compared to pure thermal relaxation at nonresonant
fields is only on the order of 25 $\%$ for $H_z=1.1$ and 2.7 kG and 75 $\%$ for $H_z=0$
(Fig.~\ref{figRelaxationDips}). This behavior is in contrast to the situation in, e.g.,
Mn$_{12}$ in absence of a PB effect, where upon QTM the relaxation time decreases by
several orders of magnitude.\cite{Thomas1996,Friedman1996} We ascribe this difference
to the fact that in Fe$_2$ the tunneling is phonon-assisted and thus slowed down by the
bottleneck effect.
\par
Importantly, our results on Fe$_2$ demonstrate that quantum tunneling of the
magnetization can be observed even though the relaxation monitored mainly represents
the phonon-bath relaxation. By creating a PB situation, QTM is observed at 1.8 K, i.e.
far above the blocking temperature for magnetization relaxation reflecting intrinsic
spin-phonon relaxation.

% Spin Reversal vs Spin Relaxation
\subsection{Modeling the Butterfly Hysteresis Loops}\label{secAvalanche}
The butterfly hysteresis reflects a spin reversal at $H_z=0$. For [Fe(salen)Cl]$_2$, it
was reproduced using a phenomenological, thermodynamic model,\cite{Shapira1999} whereas
for V$_{15}$\cite{Chiorescu2000} and Fe$_6$\cite{Waldmann2002} a dissipative two-level
model with a level anticrossing was employed. In contrast to these systems, for Fe$_2$
the presence of 11 levels and the energy barrier for spin reversal in the $S=5$ state
complicate matters considerably. As QTM is significant but not dominant in Fe$_2$
(Sec.~\ref{secQuantumtunneling}), it is likely that the spin reversal is predominantly
achieved by thermal activation over the barrier. To reproduce the butterfly shape of
the hysteresis curves in Fe$_2$, we thus decided to employ a novel, microscopic model
that explicitly accounts for the barrier. Note that although this model is based on the
absorption and emission of resonant phonons, it does not exclude the possibility of
QTM.
\par
As the resonant phonons cannot relax with the phonon bath to regain thermal
equilibrium, they produce coherent transitions of the spins on both sides of the
barrier at all values of $H_z$; however, only at $H_z=0$ these transitions lead to a
change in the macroscopically observed magnetization. In the adiabatic limit the total
energy of the combined system remains constant. Therefore the number of the absorptions
and emissions has to be equal, leading to the spin flip at $H_z=0$. The coherent spin
transitions $\Delta M=\pm1$ occur between all the spin levels $\left|M\right>$,
$M=-5,\ldots,5$. They can be mimicked mathematically by effective oscillating
transversal magnetic fields $H_{{\rm eff},x}(t)=H_{{\rm
eff},x}\sum_{-5}^{+5}\cos(\omega_{M,M+1}t)$. In its simplest form, the resulting
effective Hamiltonian for the $S=5$ state of Fe$_2$ in {\bf H} $\parallel$ {\it c}
orientation reads\cite{Leuenberger2002}
\begin{equation}
\cH_{\rm eff}=D\hat{S}_z^{2} + g\mu_BH_{{\rm eff},x}(t)\hat{S}_x + g\mu_BH_z\hat{S}_z.
\label{eqnH_eff}
\end{equation}
Following the derivation of the Hamiltonian in the generalized rotating frame
\cite{Leuenberger2004} (details see Appendix), we obtain
\begin{eqnarray}
\cH_{\rm eff}^{\rm grot} = g\mu_BH_{{\rm eff},x}\hat{S}_x + g\mu_BH_z\hat{S}_z.
\label{eqnH_grot1}
\end{eqnarray}
The levels $|m\rangle=-5,\ldots,5$ correspond to the combined spin+phonons system. Their
response $\cH_{\rm eff}^{\rm grot}$ is shown in Fig.~\ref{figLevel_anticrossing} for
$H_{{\rm eff},x}=350$ G. Within about $-0.5\leq H_z\leq+0.5$ kG, they become strongly
mixed, and at $H_z=0$ they are equidistantly split by $10g\mu_BH_{{\rm eff},x}=0.47$ K.
Note that in the generalized rotating frame the energy of a given resonant phonon is
added to the energy of a given $\left|M\right>$ level.\cite{Leuenberger2004}
Consequently, the energy difference to the adjacent level and hence the entire energy
barrier disappear. Therefore, the resonant phonons provide an efficient way to overcome
this barrier, leading to the butterfly-shaped hysteresis.
\par
%%%%%%%%%%%%%%%%%%%%%%%%%%%%%%%%%%%%%%%%%%%%%%%%%%%%%%%%%%%%%%%%%%%%%%%%%%%%%%%%%%%%%%%%%%
\begin{figure}[t!]
\includegraphics[width=75mm]{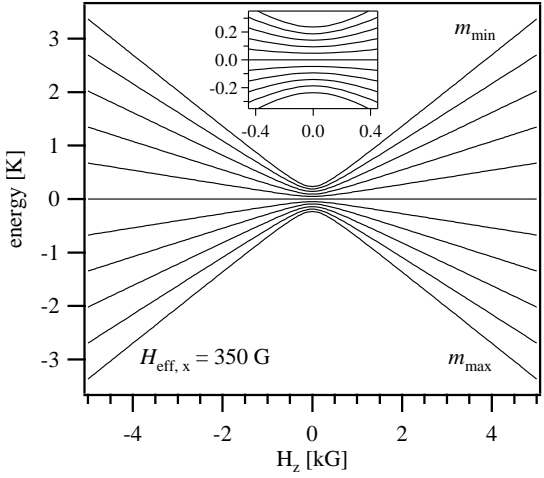}
\caption{Dependence of the 11 energy levels $|m\rangle$ defined by $\cH_{\rm eff}^{\rm
grot}$ (Eq.~\ref{eqnH_grot1}) on $H_z$, arising from the resonant phonons ($H_{\rm
eff,x}=350$ G). The inset highlights the anticrossing at $H_z=0$.}
\label{figLevel_anticrossing}
\end{figure}
%%%%%%%%%%%%%%%%%%%%%%%%%%%%%%%%%%%%%%%%%%%%%%%%%%%%%%%%%%%%%%%%%%%%%%%%%%%%%%%%%%%%%%%%%%
%
In the adiabatic limit corresponding to the data in Fig.~\ref{figAvalanche} obtained
upon a fast ($\Gamma\approx150$ G/s) sweep, the combined spin+phonons system remains
decoupled from the bath. Thus the magnetization essentially switches from positive to
negative saturation upon crossing $H_z=0$, indicating that during the entire sweep only
the lowest level in Fig.~\ref{figLevel_anticrossing} is populated. Consequently,
$M(H_z)\propto d\varepsilon_{m_{\rm max}}/dH_z$, where $\varepsilon_{m_{\rm max}}$ is
the energy of the lowest-energy level defined by $\cH_{\rm eff}^{\rm grot}$
(Fig.~\ref{figLevel_anticrossing}). With $H_{{\rm eff},x}=350$ G the agreement with the
experimental data is excellent (Fig.~\ref{figAvalanche}), thus lending credence to our
model, in which $H_{{\rm eff},x}$ is the {\it only} adjustable parameter. The range
$-0.5\leq H_z\leq+0.5$ kG in which the spin reversal occurs (Fig.~\ref{figAvalanche})
mirrors the zone in which the levels are highly mixed
(Fig.~\ref{figLevel_anticrossing}). Thus it directly depends on the magnitude of
$H_{{\rm eff},x}$ by $(dM/dH_z)_{H_z\rightarrow0}\propto1/H_{{\rm eff},x}$.
\par
In nonadiabatic situations with a slow sweeping rate, corresponding to the butterfly
hystereses in Figs.~\ref{figFullhysteresis} and \ref{figHysteresisdependence}, the
combined system is able to partially relax to thermal equilibrium during the
measurement. Thus, the fast reversal process has to be combined with the slow
relaxation process. Inside a region around $H_z=0$ we apply the resonant phonon model
({\it vide supra}) with energy levels $\varepsilon_m$ of the combined system to account
for the fast spin reversal. Additionally, the slow sweeping rate together with the weak
phonon-bath coupling gives rise to the presence of a small number of nonresonant or
thermal phonons, which allows for the slow relaxation of the combined system to the
bath. Thus, at $H_z=0$ the relaxation time becomes\cite{Leuenberger2000}
\begin{equation}
\tau_{\rm in} = \frac{1}{1+e^{(\varepsilon_{m_{\rm min}}-\varepsilon_{m_{\rm
max}})/kT}} \sum_{j=1}^{2S}\frac{e^{(\varepsilon_{m_j}-\varepsilon_{m_{\rm
max}})/kT}}{W_{m_j}^{m_{j+1}}},
\end{equation}
where $m_{\rm min}=m_1$ and $m_{\rm max}=m_{2S+1}$. The spin-thermal phonon transition
rates read
\begin{eqnarray}
W_{m_j}^{m_{j+1}} & = & \frac{g_0^2S_{m_j}^{m_{j+1}}}{48\pi\rho v^5\hbar^4}
\frac{(\varepsilon_{m_{j+1}}-\varepsilon_{m_j})^3}{e^{(\varepsilon_{m_{j+1}}-\varepsilon_{m_j})/kT}-1} \nn\\
& = & \frac{g_0^2S_{m_j}^{m_{j+1}}}{24\pi\rho v^5\hbar^4}
\frac{(\varepsilon_{m_{j+1}}-\varepsilon_{m_j})^3e^{(\varepsilon_{m_{j}}-\varepsilon_{m_{j+1}})/2kT}}
{\sinh[(\varepsilon_{m_{j+1}}-\varepsilon_{m_j})/2kT]}, \nn\\
\end{eqnarray}
where $g_0$ is the spin-thermal phonon interaction parameter, $\rho=1.36\times 10^3$
g/cm$^3$ the mass density,\cite{Schenker2001} $v$ the sound velocity, and
$S_{m_j}^{m_{j+1}}=(S-M_j)(S+M_j+1)(2M_j+1)^2$ in the quantization axis defined by
$\bH_{{\rm eff},x}$ (note that $H_z=0$).\cite{approximation} Considering only transitions
between the two lowest levels, $\tau_{\rm in}$ can be approximated by
\begin{eqnarray}
\tau_{\rm in} & \approx & \frac{1}{1+e^{(\varepsilon_{m_{\rm min}}-\varepsilon_{m_{\rm
max}})/kT}} \frac{e^{(\varepsilon_{m_{2S}}-\varepsilon_{m_{\rm
max}})/kT}}{W_{m_{2S}}^{m_{max}}} \nn\\
& = & \frac{\sinh[(\varepsilon_{m_{\rm
max}}-\varepsilon_{m_{2S}})/2kT]}{1+e^{(\varepsilon_{m_{\rm min}}-\varepsilon_{m_{\rm
max}})/kT}} \frac{e^{(\varepsilon_{m_{2S}}-\varepsilon_{m_{\rm
max}})/2kT}}{\gamma_{m_{2S}}^{m_{\rm max}}}, \nn\\
\end{eqnarray}
where $\gamma_{m_{2S}}^{m_{\rm max}}=g_0^2S_{m_{2S}}^{m_{\rm max}}(\varepsilon_{m_{\rm
max}}-\varepsilon_{m_{2S}})^3/24\pi\rho v^5\hbar^4= 810g_0^2\left(g\mu_BH_{{\rm
eff},x}\right)^3/24\pi\rho v^5\hbar^4$ at $H_z=0$.\cite{approximation} Notably, as for a
two-level system $m_{2S}=m_{\rm min}$, this equation shows the well-known $\tanh$
behavior for the relaxation time of a two-level system.\cite{Abragam1970}
\par
A general relaxation time $\tau_{\rm out}$ was chosen to account for the phonon
bottleneck effect outside the region around $H_z=0$. The time evolution of the
magnetization is given by $\dot{M}(t)=-\left[M(t)-M_{\rm eq}(H_z(t))\right]/\tau$ that
for small values of $\Gamma$ yields
\begin{equation}
M(t)=M_{\rm eq}+(M_{\rm sat}-M_{\rm eq})e^{-t/\tau}\quad,
\end{equation}
where $M_{\rm sat}=M(H_{\rm sat})$ and $M_{\rm eq}=M_{\rm eq}(H_z)$ denote the
magnetizations at saturation and thermal equilibrium, respectively. $H_z(t)=H_{\rm
sat}+\Gamma t$ was swept from $H_z=H_{\rm sat}=-10.0$ kG to $+10.0$ kG. $\tau=\tau_{\rm
in}$ for $|H_z|\le 1.5$ kG and $\tau=\tau_{\rm out}$ for $|H_z|>1.5$ kG. The value of
$H_z$ that marks the border between the two regimes is defined by the magnitude of
$H_{{\rm eff},x}$.
\par
The butterfly hysteresis in Fig.~\ref{figFullhysteresis} was best reproduced with
$\tau_{\rm in}=2100$ s, whose temperature dependence turned out to be insignificant
below 10 K. Assuming a realistic sound velocity of $v=1400$ m/s,\cite{Leuenberger2000}
this value corresponds to $g_0=0.72$ mK, i.e. about 3 orders of magnitude smaller than
the values reported for Mn$_{12}$\cite{Leuenberger2000} and
Fe$_8$\cite{LeuenbergerFe8_2000}. This low value for $g_0$ in Fe$_2$ arises from the
fact that the number of thermal phonons is very small due to the weak phonon-bath
coupling in our experiments. Hence, it does not necessarily reflect a weak spin-phonon
interaction. $\tau_{\rm out}=550$ s as obtained from the fit is in overall reasonable
agreement with the experimental values for $\tau_2$ deduced from the magnetization
relaxation curves (Fig.~\ref{figRelaxationDips}). Importantly, $\tau_{\rm in}>\tau_{\rm
out}$, which leads to the typical features of a butterfly hysteresis loop. The resonant
phonon field is $H_{{\rm eff},x}=1$ kG, which is about three times larger than the
value obtained from the analysis of the data in Fig.~\ref{figAvalanche}, indicating
that $H_{{\rm eff},x}$ depends on the sweeping rate. The best fit was obtained with
sweeping rates of 2.5 G/s for $|H_z|\le 5$ kG and 3.3 G/s for $|H_z|>5$ kG, i.e. in
very good agreement with the experimental values.\cite{sweepingrate_fnote} To fit the
butterfly hystereses at 3 and 5 K (Fig.~\ref{figHysteresisdependence}), $\tau_{\rm
out}$ and $H_{{\rm eff},x}$ were multiplied with the Boltzmann factor
$e^{\Delta(1/kT_1-1/kT_2)}$, where $T_2=1.8$ K and $T_1=3$ K or 5 K. The decrease of
$\tau_{\rm out}$ and $H_{{\rm eff},x}$ with increasing $T$ reflects the fact that (i)
the coherence and thus relevance of the resonant phonons decreases, and (ii) the
phonon-bath coupling increases.

\subsection{Intrinsic Spin-Phonon Dynamics}\label{ACanalysis}
Insight into the intrinsic spin-phonon relaxation properties of Fe$_2$ unaffected by the
PB is provided by the AC susceptibility experiments on samples with a good thermal
contact with the bath. The absence of a PB situation is confirmed by the identical
relaxation rates $1/\tau$ found for single-crystalline (sample {\bf B}) and powder
samples (sample {\bf C}) at given values of $H_{\rm DC}$, as otherwise $1/\tau$ would
increase with the size of the crystallites as phonon scattering at the boundary walls
decreases.\cite{Giordmaine1958,Scott&Jeffries1962}
\par
With an $S=5$ ground state and an energy barrier of $\Delta=2.40$ K for spin reversal,
Fe$_2$ features the principal requirements necessary for exhibiting SMM behavior.
However, from the absence of a $\chi"$ signal at $H_{\rm DC}=0$ we conclude that Fe$_2$
does not behave as an SMM at $T\geq1.8$ K. Obviously at these temperatures the
relaxation over the small barrier is beyond the detectable range; presumably the
blocking temperature for SMM relaxation lies in the mK region. However, by applying a
static field $H_{\rm DC}$, strong $\chi"$ features are readily observed. We ascribe
this difference to the fact that at $H_{\rm DC}\neq0$, phonons absorbed on one side of
the barrier can no longer resonate with those emitted on the other side, rendering the
relaxation considerably less efficient.
\par
%%%%%%%%%%%%%%%%%%%%%%%%%%%%%%%%%%%%%%%%%%%%%%%%%%%%%%%%%%%%%%%%%%%%%%%%%%%%%%%%%%%%%%%%%%%%%%%%%%%%%%%%%%%%%%%%%%%%%
\begin{figure}[!t]
\includegraphics[width=75mm]{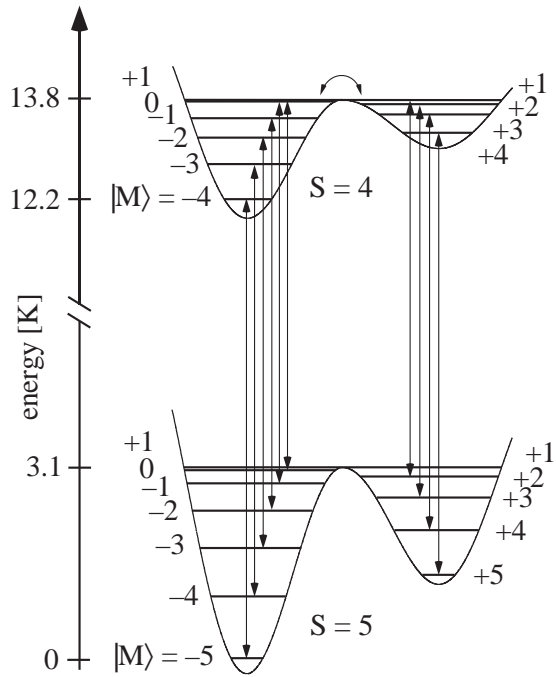}
\caption{Double well potential for a bias field of $H_{\rm DC}=1$ kG. The arrows
indicate individual $\Delta S=\Delta M\pm1$ spin-phonon transitions within the combined
$S=5$, $S=4$ spin system and illustrate a likely scenario for spin-phonon relaxation.}
\label{figPotential_Asymm}
\end{figure}
%%%%%%%%%%%%%%%%%%%%%%%%%%%%%%%%%%%%%%%%%%%%%%%%%%%%%%%%%%%%%%%%%%%%%%%%%%%%%%%%%%%%%%%%%%%%%%%%%%%%%%%%%%%%%%%%%%%%%
%
The relaxation rate $1/\tau$ is temperature dependent, indicating that the relaxation
occurs via a thermal activation process. At $H_{\rm DC}=1$ kG, the temperature
dependence of $1/\tau$ follows the Arrhenius law (Fig.~\ref{figACarrhenius}), thus
suggesting an Orbach process.\cite{Abragam1970} This behavior is typically observed for
SMMs, and often the value for $\Delta E$ is regarded as a lower limit for the SMM
barrier height (note that $\chi"$ features at $H_{\rm DC}\neq0$ have been reported for
a few SMMs including Mn$_4$,\cite{Aubin1998} V$_4$,\cite{Castro1998} and
Mn$_{12}$;\cite{Novak1995,Luis1997} for V$_4$, the $\chi"$ signal was observed only at
$H_{\rm DC}\neq0$). However, in Fe$_2$ the kinetic energy barrier of $\Delta
E=12.6\pm0.2$ K is 4 times larger than the thermodynamic barrier of the $S=5$ state at
1 kG, $\Delta=3.1$ K (Fig.~\ref{figPotential_Asymm}). Importantly, this value lies
close to the exchange splitting between the $S=5$ and the $S=4$ states of only 11.2
K\cite{Schenker2001IC} (Fig.~\ref{figPotential_Asymm} shows the anisotropy and Zeeman
splittings of these states at $H_{\rm DC}=1$ kG). This similarity hints toward a
relaxation process involving transitions to the $S=4$ state
(Fig.~\ref{figPotential_Asymm}). While the time-averaged level populations of the spin
system are defined by the Boltzmann statistics, the oscillating field induces
steady-state spin-phonon transitions on both sides of the barrier in an attempt to
achieve thermal equilibrium. At $T_{\rm max}\approx4$ K, the $S=4$ state is populated
by about 5\%, and so is the phonon energy spectrum. Assuming a Debye model, at $4$ K
the density of phonon states suitable for, e.g., an $|S,M\rangle=|5,-5\rangle$
$\longrightarrow$ $|4,-4\rangle$ transition is about two orders of magnitude higher
than that for a $|5,-5\rangle$ $\longrightarrow$ $|5,-4\rangle$ transition, resulting
in a strongly increased number of available thermal phonons for spin-phonon
transitions. Also, at 4 K all the $S=5$ levels are highly populated. Consequently, a
relaxation process involving $\Delta S=\pm1$ rather than $\Delta S=0$ transitions
(Fig.~\ref{figPotential_Asymm}) becomes favorable. Therefore we postulate that the
$S=5$ and $S=4$ states form a combined system in which the relaxation occurs from the
$S=5$ state over the barrier of the $S=4$ state. The effective energy barrier is then
13.8 K and 12.5 K for transitions from the deeper to the shallower well and {\it vice
versa}, respectively, in good agreement with the kinetic energy barrier $\Delta
E=12.6\pm0.2$ K. This agreement further indicates that $\Delta S=0$ transitions --
although still possible -- do not
play a dominant role in the relaxation process.%
%%%%%%%%%%%%%%%%%%%%%%%%%%%%%%%%%%%%%%%%%%%%%%%%%%%%%%%%%%%%%%%%%%%%%%%%%%%%%%%%%%%%%%%%%%%%%%%%%%%%%%%%%%%%%%%%%%%%%
\begin{figure}[htb]
\includegraphics[width=75mm]{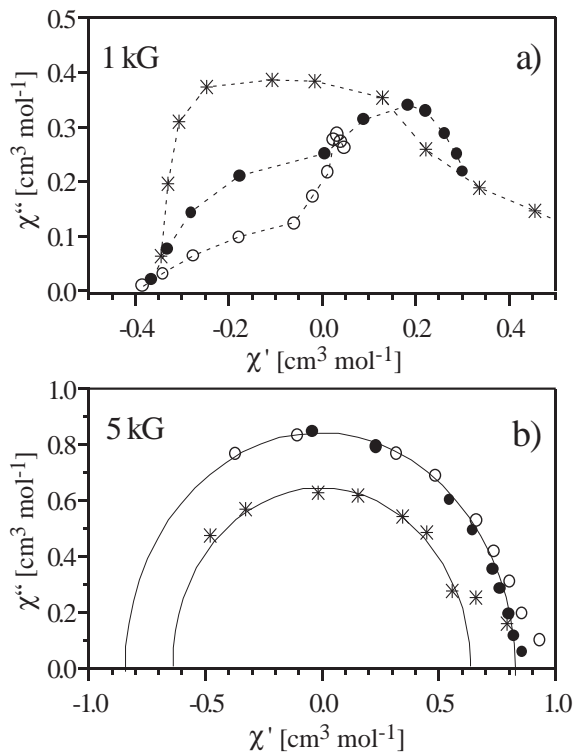}
\caption{Argand or Cole-Cole plots of $\chi$" vs $\chi$' at $H_{\rm DC}=1$ (a) and 5 kG
(b) at fixed temperatures ($\Large{\ast}$, 3.0 K; $\Large{\bullet}$, 4.0 K;
$\Large{\circ}$, 5.0 K). For clarity the data are shifted along the x axis. For a
single relaxation process the data points should lie on a semicircle (solid lines) at
each temperature.} \label{figACColeCole}
\end{figure}
%%%%%%%%%%%%%%%%%%%%%%%%%%%%%%%%%%%%%%%%%%%%%%%%%%%%%%%%%%%%%%%%%%%%%%%%%%%%%%%%%%%%%%%%%%%%%%%%%%%%%%%%%%%%%%%%%%%%%
\par
Fig.~\ref{figACColeCole}a shows the so-called Cole-Cole or Argand plot for the AC
susceptibility data at $H_{\rm DC}=1$ kG, in which $\chi$" is plotted vs $\chi$' for
all frequencies at fixed temperatures as indicated. Importantly, for a given
temperature the data do not lie on a semicircle but deviate
widely,\cite{ACcolecole_fnote} thus unambiguously indicating the presence of a
multitude of relaxation processes with different rates. These rates can be calculated
using Pauli's master equation in conjunction with Fermi's golden rule, and are obtained
as the eigenvalues of the matrix of transition
rates.\cite{Leuenberger2000,Villain1994,Hartmann1996,Fort1998,Fernandez1998} The
smallest nonzero eigenrate corresponds to the rate at which the relative populations of
both wells reach mutual equilibrium, i.e. to the relaxation of the spin system over the
barrier. All the faster rates are associated with transitions between levels inside
each of the two wells. In Mn$_{12}$, e.g., the smallest relaxation rate is about four
orders of magnitude smaller than the second-smallest.\cite{Leuenberger2000}
Consequently, Mn$_{12}$ relaxes with a single observable relaxation rate. Obviously,
for Fe$_2$ the situation is different. The small energetic spread of individual
spin-phonon transitions leads to a very narrow distribution of eigenrates that are not
resolved experimentally. Therefore, the relaxation phenomenon observed in the AC
susceptibility measurements of Fe$_2$ does not exclusively correspond to the relaxation
over the $S=5/S=4$ effective energy barrier but also reflects individual $\Delta S\pm1$
spin-phonon transitions on either side of this barrier.
\par
At $H_{\rm DC}=5$ kG the situation is distinctly different. The $1/\tau\propto T$
dependence (Fig.~\ref{figACarrhenius}) is consistent with a direct rather than an
Orbach process, with a spin-phonon transition energy $\hbar\omega< T=3-5$
K.\cite{Abragam1970} Also, the Argand plot (Fig.~\ref{figACColeCole}b) indicates that
$1/\tau$ essentially consists of a single relaxation rate. These results suggest that
at 5 kG, the observed relaxation is governed by the transitions between two levels
only, which, however, are difficult to identify. Nevertheless, a relaxation process
involving the $S=4$ excited state appears highly unlikely, because (i)  at 5 kG the
$S=4$ state has lost its barrier entirely, and (ii) $\hbar\omega\ll T=3-5$ K.
Consequently, the monitored relaxation is governed by transitions between two levels of
the $S=5$ state.
\par
The observed increase of $\chi"$ between 1 and 5 kG (Fig.~\ref{figACfrequency},
Sec.~\ref{secACresults}) indicates that the relaxation process becomes increasingly
inefficient with $H_{\rm DC}$. A number of reasons may contribute to this behavior.
Perhaps the most important one is that the number of possible spin-phonon transitions
involved in the relaxation process shrinks markedly with increasing asymmetry of the
potential well, concomitant with a change from an Orbach to a direct relaxation
process.

\section{Summary and Conclusions}\label{secConclusions}
This study provides significant insight into the unusual dynamics of the magnetic
properties of the spin cluster [Fe$_2$F$_9$]$^{3-}$.
\par
First, the observed slow magnetization relaxation in Fe$_2$ arises from poor thermal
contact between the sample and the bath. The resulting phonon bottleneck situation allows
for the indirect observation of quantum tunneling of the magnetization through the energy
barrier at 1.8 K. Importantly, this result illustrates that a PB situation can provide
insight into the dynamics of the spin system at temperatures far above the blocking
temperature below which spin-phonon relaxation becomes directly detectable.
\par
Second, in the high-spin system Fe$_2$ the hysteresis caused by the PB effect adopts a
butterfly shape that is phenomenologically analogous to that observed for the low-spin
systems V$_{15}$ and Fe$_6$. This result indicates that despite the energy barrier in
Fe$_2$, the spin reversal at $H_z=0$ occurs much faster than the re-equilibration of
the spin+phonons system with the bath. As QTM is rather inefficient in Fe$_2$, it
further suggests that thermal activation is competitive in achieving spin reversal. Our
microscopic model based on the rapid absorption/emission of resonant phonons allows for
an accurate reproduction of the hysteresis curves observed for Fe$_2$.
\par
Third, AC susceptibility experiments unobstructed by the phonon bottleneck allow for
direct insight into spin-phonon dynamics. Even on the timescale of our AC experiments,
Fe$_2$ does not behave as an SMM at $T\geq1.8$ K. At $H_{\rm DC}=1$ kG an out-of-phase
susceptibility is detected with a frequency dependence closely resembling the one
typically observed for SMMs; hovever, the situation in Fe$_2$ is substantially more
complex and highly unusual. The $S=5$ ground and the $S=4$ first excited states form a
combined system in which the relaxation occurs over the barrier of the combined system,
favored by the much higher density of states for thermal phonons suitable for $\Delta
S=\pm1$ than $\Delta S=0$ transitions. Importantly, this result demonstrates that the
measured kinetic energy barrier can actually be {\it larger} than the thermodynamic
barrier of the ground state if the excited state becomes thermally accessible. This
result provides an additional twist for the interpretation of the kinetic energy
barrier of single-molecule magnets derived from AC susceptibility measurements.

\begin{acknowledgments}
The authors thank Stefan Ochsenbein for his aid in performing some of the magnetic
measurements, and Dr. Oliver Waldmann for insightful discussions. This work has been
supported by the Swiss National Science Foundation and by the TMR program Quemolna of
the EU.
\end{acknowledgments}

% APPENDIX
\appendix*
\section{}\label{secAppendix}
Eq.~(\ref{eqnH_grot1}) is derived from Eq.~(\ref{eqnH_eff}) as follows. Generalizing
the rotating wave approximation,\cite{Sakurai} the Hamiltonian in Eq.~(\ref{eqnH_eff})
can be approximated by
\begin{equation}
\cH_{\rm eff}\approx\left[\begin{array}{ccccc}
25D & h_{5,4} & 0 & \cdots & 0 \\
h_{4,5} & 16D & h_{4,3} & \ddots & \vdots \\
0 & h_{3,4} & 9D & \ddots & 0  \\
\vdots & \ddots & \ddots & \ddots & h_{-4,-5} \\
0 & \cdots & 0 & h_{-5,-4} & 25D \ea + g\mu_BH_z\hat{S}_z, \label{eqnH_approx}
\end{equation}
where $h_{M+1,M}=g\mu_BH_{{\rm eff},x}\sqrt{(S-M)(S+M+1)}\\ e^{i\omega_{M+1,M}t}/2$ and
$h_{M,M+1}=h_{M+1,M}^*$. To remove the time dependence of Hamiltonian
(\ref{eqnH_approx}), a unitary transformation $U$ is applied to
Eq.~(\ref{eqnH_approx}). In order to obtain the transformation of $\cH_{\rm eff}$ to
the generalized rotating frame, $U$ has to be applied to the time-dependent
Schr\"odinger equation
\begin{equation}
i\hbar\frac{\p\left|\psi\right>}{\p t}=\cH_{\rm eff}\left|\psi\right>.
\label{eqnSchrödinger}
\end{equation}
We define the quantum state in the generalized rotating frame by $\left|\psi_{\rm
grot}\right>=U\left|\psi\right>$. The transformation of the left and right hand side of
Eq.~(\ref{eqnSchrödinger}) then yields
\begin{eqnarray}
i\hbar\frac{\p\left|\psi\right>}{\p t} & = & i\hbar\left(
U^{-1}\frac{\p\left|\psi_{\rm grot}\right>}{\p t}+\frac{\p U^{-1}}{\p t}\left|\psi_{\rm grot}\right>\right), \nn\\
\cH_{\rm eff}\left|\psi\right> & = & \cH_{\rm eff} U^{-1}\left|\psi_{\rm grot}\right>.
\label{eqnSchrödinger2}
\end{eqnarray}
Combining the left and right hand side of Eq.~(\ref{eqnSchrödinger2}) leads to
\begin{equation}i\hbar\left( U^{-1}\frac{\p\left|\psi_{\rm grot}\right>}{\p t}+\frac{\p U^{-1}}{\p
t}\left|\psi_{\rm grot}\right>\right) =\cH_{\rm eff} U^{-1}\left|\psi_{\rm
grot}\right>. \label{eqnCombination}
 \end{equation}
Multiplying Eq.~(\ref{eqnCombination}) by $U$ from the left results in
\begin{eqnarray}
i\hbar\frac{\p\left|\psi_{\rm grot}\right>}{\p t} & = & \left(U\cH_{\rm eff}
U^{-1}-i\hbar U\frac{\p U^{-1}}{\p t}\right)
\left|\psi_{\rm grot}\right> \nn\\
& \equiv & \cH_{\rm eff}^{\rm grot}\left|\psi_{\rm grot}\right>,
\end{eqnarray}
from which we can directly read off the transformed Hamiltonian
\begin{equation}
\cH_{\rm eff}^{\rm grot}=U\cH_{\rm eff} U\D-i\hbar U\frac{\p U\D}{\p t}=U\cH_{\rm eff}
U\D+i\hbar\frac{\p U}{\p t}U\D.
\end{equation}
We now choose the following unitary transformation:
\begin{equation}
U=\sum_{M=-5}^5\left|M\right>\left<M\right|e^{-i\omega_Mt}.
\end{equation}
Evaluating $\cH_{\rm eff}'=U\cH_{\rm eff}U\D$ yields
\begin{equation}
\cH_{\rm eff}'=\left[\begin{array}{ccccc}
25D & h_{5,4}' & 0 & \cdots & 0 \\
h_{4,5}' & 16D & h_{4,3}' & \ddots & \vdots \\
0 & h_{3,4}' & 9D & \ddots & 0  \\
\vdots & \ddots & \ddots & \ddots & h_{-4,-5}' \\
0 & \cdots & 0 & h_{-5,-4}' & 25D \ea + g\mu_BH_z\hat{S}_z,
\end{equation}
where $h_{M+1,M}'=h_{M+1,M}e^{i(\omega_M-\omega_{M+1})t}$ and
$h_{M,M+1}'=h_{M+1,M}'^*$. Setting $\omega_{M+1,M}-\omega_{M+1}+\omega_M=0$ eliminates
the time dependence of
\begin{eqnarray}
\cH_{\rm eff}^{\rm grot} & = & \cH_{\rm eff}'+i\hbar\frac{\p U}{\p t}U\D \nn\\
& = & g\mu_BH_{{\rm eff},x}\hat{S}_x + g\mu_BH_z\hat{S}_z. \label{eqnH_grot}
\end{eqnarray}

% REFERENCES


\begin{thebibliography}{}
\bibitem{sweepingrate_fnote}As our SQUID magnetometer does not allow for continuous
detection of the magnetic moment upon sweeping the field, the reported sweeping rates
$\Gamma$ are time-averaged and depend on the number of data points within a certain
magnetic field interval. The value of $\Gamma\approx 3$ G/s reported for the butterfly
loops actually reads $\Gamma=2.5$ G/s for $0\leq\left|H_z\right|\leq5$ kG and 3.5 G/s
for $\left|H_z\right|>5$ kG.
\bibitem{ACXtalpowder_fnote}The [Fe$_2$F$_9$]$^{3-}$ dimer molecules are
aligned along $\bH_{\rm AC}\parallel\bH_{\rm DC}\parallel c$ in the crystal, while they
are isotropically oriented in the polycrystalline sample. This difference results in a
much larger number of spins not being able to adiabatically follow the oscillating field
in the single crystal, leading to a more intense and sharper $\chi$" signal.
\bibitem{approximation} Note that the definition of $\tau_{\rm in}$ is only strictly valid at
$H_z=0$. However, it remains a valid approximation for small magnitudes of $H_z$.
\bibitem{ACcolecole_fnote} Varying contributions from $\Delta S=0$ and $\Delta S=\pm1$
transitions may have a significant impact on the appearance of the Argand diagram.
Notably, while the Arrhenius analysis focuses on the situation at the maximum of the
$\chi"$ profiles in Fig.~\ref{figACfrequency}, the Argand diagram plots also $\chi"$
values from the wings of these profiles.
\end{thebibliography}
\end{document}